\documentstyle [11pt]{article}
\def\journal#1#2#3#4{\  {#1} \ {\bf #2}, {#3}\  ({#4})}
\def\NPB{\journal{Nucl.\ Phys.\ {\bf B}}}

\def\PhysRev{\journal{Phys.\ Rev.}}

\begin{document}
\pagestyle{empty}

\thispagestyle{empty}

\newcommand{\gn}{\mbox{$\gamma_{\stackrel{}{5}}$}}
\newcommand{\adag}{a^{\dagger}_{p,s}}
\newcommand{\atildedag}{\tilde{a}^{\dagger}_{-p,s}}
\newcommand{\bdag}{b^{\dagger}_{-p,s}}
\newcommand{\btildedag}{\tilde{b}^{\dagger}_{-p,s}}
\newcommand{\apsbeta}{a^{\beta}_{p,s}}
\newcommand{\apsbetadag}{a_{-p,s}^{\beta\dagger}}
\newcommand{\adagbdag}{a^{\dagger}_{p,s} b^{\dagger}_{-p,s}}
\newcommand{\aps}{a^{}_{p,s}}
\newcommand{\bps}{b^{}_{-p,s}}
\newcommand{\bpsbeta}{b^{\beta}_{p,s}}
\newcommand{\bpsbetadag}{b_{-p,s}^{\beta\dagger}}
\newcommand{\Adag}{A^{\dagger}_{p,s}}
\newcommand{\Bdag}{B^{\dagger}_{-p,s}}
\newcommand{\Aps}{A^{}_{p,s}}
\newcommand{\Apsbeta}{A^{\beta}_{p,s}}
\newcommand{\Apsbetadag}{A^{\beta\dagger}_{-p,s}}
\newcommand{\Bps}{B^{}_{p,s}}
\newcommand{\Bpsbeta}{B^{\beta}_{p,s}}
\newcommand{\Bpsbetadag}{B^{\beta\dagger}_{-p,s}}
\newcommand{\ApL}{A^{}_{p,L}}
\newcommand{\BpL}{B^{}_{-p,L}}
\newcommand{\ApR}{A^{}_{p,R}}
\newcommand{\BpR}{B^{}_{-p,R}}
\newcommand{\apL}{a^{}_{p,L}}
\newcommand{\bpL}{b^{}_{-p,L}}
\newcommand{\apR}{a^{}_{p,R}}
\newcommand{\bpR}{b^{}_{-p,R}}
\newcommand{\AdagL}{A^{\dagger}_{p,L}}
\newcommand{\AdagR}{A^{\dagger}_{p,R}}
\newcommand{\BdagL}{B^{\dagger}_{-p,L}}
\newcommand{\BdagR}{B^{\dagger}_{-p,R}}
\newcommand{\adagL}{a^{\dagger}_{p,L}}
\newcommand{\adagR}{a^{\dagger}_{p,R}}
\newcommand{\bdagL}{b^{\dagger}_{-p,L}}
\newcommand{\bdagR}{b^{\dagger}_{-p,R}}
\newcommand{\eps}{\epsilon}
\newcommand{\gnplus}{\gamma \cdot n_{_{+}}}
\newcommand{\gnminus}{\gamma \cdot n_{_{-}}}
\newcommand{\gnplusdef}{\left( \vec{\gamma} \cdot \hat{n}-\gamma_o \right)}
\newcommand{\gnminusdef}{\left( \vec{\gamma} \cdot \hat{n}+\gamma_o \right)}
\newcommand{\abab}{a^{\dagger}_{p,L}\,b^{\dagger}_{-p,L}\,a^{\dagger}_{p,R}
                   \,b^{\dagger}_{-p,R}}
\newcommand{\alphai}{\alpha_{i}}
\newcommand{\limit}{\lim_{\Lambda^2 \rightarrow \infty}}
\newcommand{\p}{\vec{p}, p_o}
\newcommand{\poprime}{p_o^{\prime}}
\newcommand{\prodps}{\prod_{p,s}}
\newcommand{\prodp}{\prod_{p}}
\newcommand{\psibar}{\bar{\psi}}
\newcommand{\psibarpsi}{ < \bar{\psi} \, \psi
            > }
\newcommand{\PsibarPsi}{ < \bar{\Psi} \, \Psi
            > }
\newcommand{\psibeta}{\psi^{}_{\beta}}
\newcommand{\psibarbeta}{\bar{\psi}_{\beta}}
\newcommand{\psidag}{\psi^{\dagger}}
\newcommand{\psidagbeta}{\psi^{\dagger}_{\beta}}
\newcommand{\psiL}{\psi_{_{L}}}
\newcommand{\psiR}{\psi_{_{R}}}
\newcommand{\Q}{Q_{_{5}}}
\newcommand{\Qa}{Q_{_{5}}^{a}}
\newcommand{\Qbeta}{Q_{5}^{\beta}}
\newcommand{\qqbar}{q\bar{q}}
\newcommand{\sumps}{\sum_{p,s}}
\newcommand{\thetap}{\theta_{p}}
\newcommand{\costhetap}{\cos{\thetap}}
\newcommand{\sinthetap}{\sin{\thetap}}
\newcommand{\thetaset}{\{ \thetap  \}}
\newcommand{\thetapi}{\thetap{}_{i}}
\newcommand{\Tomega}{\frac{\Tprime}{\omega}}
\newcommand{\pomega}{\frac{p}{\omega}}
\newcommand{\Tprime}{T'}
\newcommand{\Tprimesq}{T^{'2}}
\newcommand{\vac}{| vac \rangle}
\newcommand{\vacbeta}{| vac \rangle_{_{\beta}}}
\newcommand{\x}{\vec{x},t}
\newcommand{\xPrime}{\vec{x} - \hat{n} (t - t' ), t'}
\newcommand{\xPrimet}{\vec{x} + \hat{n} (t - t'), t'}
\newcommand{\y}{\vec{y}, y_o}

\setlength{\unitlength}{1in}
\begin{picture}(3,0)
\put(4.5,.25){\framebox(2,.25){CCNY-HEP-94-9}}
\end{picture}

\begin{center}
{\Large {\bf
Chiral Restoration in the Early Universe: \\
Pion Halo in the Sky
}
%\fnsymbol{footnote}\footnote
%{\samepage \sl
%\noindent \parbox[t]{6in}{\noindent
%                         }
%}\\
%{\bf }
}\\
\end{center}
\baselineskip .2in
\begin{center}
Ngee-Pong Chang (npccc@cunyvm.cuny.edu)\\
Department of Physics\\
City College \& The Graduate School of City University of New York\\
New York, N.Y. 10031\\
\ \\
September 28, 1994 \\
\end{center}

%\begin{center}
%{\bf Abstract} \\
%\ \\
%\parbox[t]{5in}{\small \sl
%
%
%
%               }
%\end{center}

\section{Introduction}

        I am very pleased to have the opportunity to present to this
        distinguished audience some recent developments concerning chiral
        symmetry of the early universe.  Chiral restoration is so taken for
        granted that it has not even been raised by others at this
        astroparticle workshop.

        As you will see, there is indeed a chiral symmetry at high $T$,
        but this `restored' chirality is a morphosis of the old zero
        temperature chirality.  The original NJL vacuum undergoes an
        interesting {\em  new phase transformation} such that $\psibarpsi$
        vanishes, but the vacuum continues to break our zero temperature
        chirality.  The pion remains a Nambu-Goldstone boson,
        and actually acquires a halo while propagating through the early
        universe.

        The pion has always played a ubiquitous role in strong interaction
        physics.  In the conventional scenario, however, it has not been given
        any role at high $T$ but is ignominiously dismissed in the early
        universe, and condemned to dissociate in the early alphabet soup.
        The results reported here correctly restore the
        pion to its rightful place in the early universe.

        The pion is a messenger of an underlying broken symmetry of the
        universe, {\em  viz.} that of chirality, under the transformation
%\begin{equation}
        $\psi (\x)  \rightarrow  {\rm e}^{i \alpha \gn} \; \psi (\x)$.
%\end{equation}
        The chiral charge, $\Q$, which generates this transformation
\begin{equation}
        \Q = \int d^3 x  \;\psi^{\dagger} (\x)
                    \gn \psi (\x)               \label{eq-old-Q}
\end{equation}
        does not annihilate the vacuum.  Instead, acting on the NJL
        vacuum$\cite{NJL}$, it generates, up to a normalization factor, the
        state for a {\em  zero momentum pion},
%\begin{equation}
        $ \;\Q |vac> \;\propto\;  | \vec{\pi} (\vec{p} = 0 ) \rangle   $,
%\end{equation}
        where ($s= \mp 1$ for $L,R$ helicities)
\begin{equation}
        | vac > \;=\; \prod_{p,s} \left( \costhetap \;-\; s \,
                        \sinthetap \, \adagbdag
                        \right) \; | 0 >	\label{eq-NJL-vac}
\end{equation}
        Using the fact that $\Q$ is a constant of motion, it is easy to
        show directly that this zero momentum pion, $\Q |vac>$,
        has zero energy, thus confirming the status of the pion as a QCD
        Nambu-Goldstone boson$\cite{Goldstone}$.

        A signature of this dynamical symmetry breaking is the familiar order
        parameter, $\psibarpsi$.
        For $T > T_c$, however, it is well known that
        $\psibarpsi$ vanishes.  Chiral symmetry is said to be restored
        at $T_c$, but is it the {\em  same old chiral symmetry we knew
        at $T=0$ ?}

\section{High Temperature Effective Action}
        At high temperatures, lattice work as well as continuum field theory
        calculations show that the effective action indeed exhibits a manifest
        chiral symmetry. In thermal field theory, there is
        the famous BPFTW action$\cite{BP}$ that describes the
        propagation of a QCD fermion through a hot medium
        ($T^{'2}  \equiv \frac{\textstyle g_r^2 }{\textstyle 3} T^2$,
        while the angular brackets denote an average over the orientation
        $\hat{n}$)
\begin{equation}
   {\cal L}_{\rm eff} = - \psibar \gamma_{\mu} \partial^{\mu}
                          \psi
                       - \frac{T^{'2}}{2\;\;} \, \psibar
                         \left<
                       \frac{\gamma_o - \vec{\gamma} \cdot \hat{n} }
                       {D_o + \hat{n} \cdot \vec{D} }
                         \right> \psi          		\label{eq-BP-action}
\end{equation}
        and we see the global chiral symmetry of the action.  But the
        {\em  nonlocality} of the action implies that the Noether charge
        for this new chirality is not the same as that in eq.(\ref{eq-old-Q}).

        The fermion propagator that results from this action shows a
        pseudo-Lorentz invariant particle pole of mass $\Tprime$
        (the so-called thermal mass).  But, in addition, there is
        a pair of conjugate {\em  spacelike} plasmon cuts in the
        $p_o$-plane that run just above and below the real
        axis$\cite{Chang-xc}$, from $p_o = -p$ to $p_o = p$.  As a result,
        for $t>0$, say, the propagator function takes the form
\begin{eqnarray}
        < T( \psi(x) \bar{\psi}(0) ) >_{_{\beta}} &=& < \psi(x)
             \bar{\psi} (0) >  \\
             &=&     \;\;\; \int \frac{d^3 p}{ (2\pi)^3 } \;
                     {\rm e}^{i \vec{p} \cdot \vec{x}} \;\left\{
                     Z_{p} \frac{-i \vec{\gamma}
                     \cdot \vec{p} + i \gamma_o \omega }{2 \omega} \;
                     {\rm e}^{ - i \omega t} \right.\\
             & &     - \left. \frac{\Tprimesq}{8\;\;} \;
                     \int_{-p}^{p} \,\frac{dp_o'}{p^3}  \;
                     \frac{i \vec{\gamma} \cdot \vec{p} p_o'
                     - i \gamma_o p^2}{p^2 - p_o^2 + \Tprimesq}   \;
                     {\rm e}^{- i p_o' t} \right\}
                     + O (T'^4)                 \label{eq-spacelike-cut}
\end{eqnarray}

        In a recent study of the spacetime quantization of the
        BPFTW action$\cite{Chang-bp-local}$, I have shown that
        the spacelike cuts dictate a new thermal vacuum of the
        type
\begin{equation}
        | vac' > \;=\; \prod_{p,s} \left( \costhetap \;-\; i\, s \,
                                   \sinthetap \, \adagbdag
                       \right) \; | 0 >         \label{eq-new-vac}
\end{equation}
        The $90^{o}$ phase here in the generalized NJL vacuum is
        the reason why $\psibarpsi$ vanishes for $T \geq
        T_c$.

        The quantization of a nonlocal action is of course a technical
        matter.  Suffice it here to say that the quantization has been
        formulated in terms of auxiliary fields so that the resulting action
        is local. In this context, the pseudo-Lorentz particle pole is
        described in terms of the massive canonical Dirac field, $\Psi$,
        and the spacelike cuts are associated with the auxiliary fields, which
        are functions of $\Psi$.
        This formulation allows for a systematic expansion
        of the $\psi$ field in terms of the massive canonical Dirac field,
        $\Psi$.
        Let the $t=0$ expansion for the original massless $\psi$ field read
\begin{equation}
         \psi (\vec{x}, 0) = \frac{1}{\sqrt{V}}  \sum_{p}
                \; {\rm e}^{i \vec{p} \cdot \vec{x}}
                \left( \begin{array}{l}
                \chi_{_{p,L}} a^{}_{p,L} \;+\;
                \chi_{_{p,R}} b^{\dagger}_{-p,R}  \\
                \chi_{_{p,R}} a^{}_{p,R} \;-\;
                \chi_{_{p,L}} b^{\dagger}_{-p,L}
                       \end{array} \right)
\end{equation}
        with a corresponding canonical expansion for the massive $\Psi$,
        then we find
\begin{eqnarray}
        \aps	&=& \Aps \;-\; i \;s\; \frac{\Tprime}{2 p}
                    \Bdag + O(\Tprime {}^2 )	\label{eq-aps-Aps-1} \\
                b^{}_{p,s} &=& B^{}_{p,s} \;+\; i \;s\; \frac{\Tprime}{2 p}
                    A^{\dagger}_{-p,s}
                    + O(\Tprime {}^2)	        \label{eq-bps-Bps-1}
\end{eqnarray}
        The $O(\Tprime)$ terms in the Bogoliubov transformation
        imply the new thermal vacuum of eq.(\ref{eq-new-vac}).

        The chiral charge at high $T$ is given by
\begin{equation}
        Q_{5}^{\beta} =  - \frac{1}{2} \; \sum_{p,s}\; s \;
                      \left(
                      A^{\dagger}_{p,s} A^{}_{p,s} + B^{\dagger}_{-p,s}
                      B^{}_{p,s}
                                  \right)
\end{equation}
        so that it clearly annihilates the new thermal vacuum,
        in direct contrast with the $T=0$ Noether charge
\begin{equation}
        Q_{_{5}}
        = - \frac{1}{2}
             \sum_{p,s}\, s \; \left( a^{\dagger}_{p,s} a^{}_{p,s} +
             b^{\dagger}_{-p,s} b^{}_{-p,s}
                               \right)          \label{Q5}
\end{equation}
        which clearly fails to annihilate the vacuum at high $T$.

\section{$\psibarpsi$ is an Incomplete Order Parameter}

        The traditional order parameter $\psibarpsi$
        cannot by itself give a full description of the nature
        of chiral symmetry breaking.  The operator, $\bar{\psi} \psi$,
        belongs to a non-Abelian chirality algebra$\cite{Chang-chiralg}$,
        $SU(2N_f)_{p} \otimes SU(2N_f)_{p}$.
        The original chiral broken ground state may be written as
        $ |vac> = \prod_{p} {\rm e}^{i X_{2p} \thetap}\; |0>$,
        where $X_{2p}$ is an element of the algebra, while
        the new thermal vacuum is generated by a different element,
        $Y_{2p}$.

        Our results here suggest the study of a new class of nonlocal
        order parameters,
\begin{equation}
        - \frac{i}{\pi}  \int d^3 x \int_{-\infty}^{\infty} dt'
                  < \bar{\psi} (\x) \psi( \vec{x}, t') > + c.c.
\end{equation}
        which if nonvanishing would indicate the
        continued breaking of chiral symmetry.
        The integration over $t'$ projects away the usual timelike
        spectrum of the operator $\psi$, and probes directly the
        properties of the spacelike cut.  In our perturbative study
        here, this order parameter indeed is nonvanishing, being given by
        $ - 2 \sum_{p} \frac{\Tprime}{p^2}$, analogous to the
        familiar expression for $\psibarpsi$ at $T=0$, given by
        $ - 2 \sum_{p} \frac{M}{\sqrt{p^2 + M^2}}$, where $M$ is the
        mass gap parameter.

\section{Pion halo in the Sky}

        The pion we know at zero temperature is not massless,
        but has a mass of $135 \;MeV$.  This is because of
        electroweak breakdown, giving rise to a primordial quark mass at
        the tree level.
        At very high $T$, when electroweak symmetry is restored, we
        have the interesting new possibility that the pion will fully
        manifest its Nambu-Goldstone nature and remain physically
        massless.$\cite{Chang-QCD}$

        The pion is described by an interpolating field operator,
%\begin{equation}
         $ \sim   i \bar{\psi} \gn T^{a} \; \psi$,
%\end{equation}
        which does not know about temperature. It is the vacuum that
        depends on $T$.
        The state vector for a zero momentum pion at high $T$ may
        be obtained from the thermal vacuum by the action
%\begin{equation}
        $\Q^{a} | vac' > \;\propto \; | \pi^{a} ( \vec{p} = 0 ) \rangle   $.
%\end{equation}
        This pion now has the property that even though it is
        massless, it can acquire a {\em  screening mass} proportional
        to $T$.
        This is the pion mass that has
        been measured on the lattice at high $T$.

        As a result, the pion propagates in the early universe with a
        halo.  The retarded function for the pion shows that the
        signal propagates along the light cone, with an additional
        exponentially damped component coming from the past history
        of the source.
\begin{equation}
        D_{\rm ret} (\x) = \theta (-t) \left\{ \delta(t^2 - r^2)
                    + \frac{\Tprime}{r} \theta(t^2 - r^2)
                    \left[ {\rm e}^{-\Tprime | t-r| }
                    +  {\rm e}^{-\Tprime | t+r| } \right] \right\}
\end{equation}
        The screening mass leads to an accompanying modulator
        signal that `hugs' the light cone, with a screening length
        $\propto 1/T$.

        What are the cosmological consequences of a pion in the
        alphabet soup of the early universe?

        I am not an expert, and part of my purpose in coming to this
        workshop is to learn from you.  But one thing I know.
        In the usual scenario, the pion after chiral restoration
        will have acquired mass $\propto T$, and will quickly dissociate
        into constituent quark-antiquark pair.  According to our
        new understanding, however, the Nambu-Goldstone theorem forces
        the pion to remain a strictly massless bound state at high $T$,
        and so the pion will contribute to the partition function
        of the early universe.

        Fortunately, the pion does not contribute so many degrees of freedom
        as to upset the usual picture of the cooling of the universe.  But
        I leave it to experts to help figure out the subtle changes there
        must surely be in the phase transitions of the early universe.

        In the beginning there was light, and quarks, and gluons, to which
        we must now add the pions with halo.


\begin{thebibliography}{99}


\bibitem{NJL} Y. Nambu and G. Jona-Lasinio,  Phys. Rev.\,
     {\bf 122}, 345 (1961); {\em ibid}\ {\bf 124}, 246
     (1961).
\bibitem{Goldstone}   J. Goldstone, Nuovo Cimento {\bf 19 },
                      154 (1961).
\bibitem{BP}
        J.C.Taylor and S.M.H.Wong, \NPB{346}{115}{1990};
        E. Braaten and R. Pisarski, \PhysRev {D45}{1827}{1992};
        J. Frenkel and J.C. Taylor, \NPB{374}{156}{1992}.
\bibitem{Chang-xc}
        H.A. Weldon, Phys. Rev. \, {\bf   D40}, 2410 (1989);
        N.P. Chang, Phys. Rev. \, {\bf D 50}, 5403 (1994).
\bibitem{Chang-bp-local}
        N.P. Chang, {\em  Spacetime Quantization of BPFTW Action:
        Spacelike Plasmon Cut \& New Phase of the Thermal Vacuum },
        CCNY-HEP-94-8, Sept 21, 1994.
\bibitem{Chang-chiralg}
        N.P. Chang, {\em  Chirality Algebra}, CCNY-HEP-94-9, Sept 28, 1994.

\bibitem{Chang-QCD}
        L.N. Chang, N.P. Chang, \PhysRev {D45}{2988} {1992}.


\end{thebibliography}
\end{document}